\documentclass[%
 reprint,
superscriptaddress,
 aps,
prl,
]{revtex4-1}

\usepackage{graphicx}
\usepackage{dcolumn}
\usepackage{bm}
\usepackage{color}
\usepackage{amsmath}
\usepackage{amssymb}

\begin{document}


\title{Interplay Between Phonons and Anisotropic Elasticity Drives Negative Thermal Expansion in PbTiO$_3$}%

\author{Ethan T. Ritz}
 \email{etr38@cornell.edu}
 \affiliation{Sibley School of Mechanical and Aerospace Engineering, Cornell University, Ithaca, NY 14853 USA}
\author{Nicole A. Benedek}%
 \email{nbenedek@cornell.edu}
 \affiliation{Department of Materials Science and Engineering, Cornell University, Ithaca, NY 14853 USA}

\date{\today}

\begin{abstract}
We use first-principles theory to show that the ingredients assumed to be essential to the occurrence of negative thermal expansion (NTE) -- rigid unit phonon modes with negative Gr\"{u}neisen parameters -- are neither sufficient nor necessary for a material to undergo NTE. Instead, we find that NTE in PbTiO$_3$ involves a delicate interplay between the phonon properties of a material (Gr\"{u}neisen parameters) and its anisotropic elasticity. These unique insights open new avenues in our fundamental understanding of the thermal properties of materials, and in the search for NTE in new materials classes.

\end{abstract}

\pacs{Valid PACS appear here}
\maketitle





The first report of negative thermal expansion (NTE) in solids appeared in 1907, with Scheel's observations that quartz and vitreous silica shrink upon heating \cite{scheel1907ueber,scheel1907versuche}. 
NTE has since been observed in essentially every materials class, including metals \cite{guillaume1897recherches}, polymers \cite{bruno1998thermal}, metal-organic frameworks \cite{goodwin2008colossal, dubbeldam2007exceptional, han2007metal}, and semiconductors \cite{jiang2004thermal, biernacki1989negative, dolling1966thermodynamic}. However, despite over a century of study \cite{barrera2005negative,lind2012two,barron1980thermal}, the microscopic mechanisms of NTE remain poorly understood in all but a handful of cases \cite{tucker2005negative, van1999origin}.

In inorganic framework materials, so-called rigid unit phonon modes (RUMs) \cite{pryde1997rigid,miller2009negative} are widely recognized as being important drivers of NTE. Rather than longitudinal stretching of bonds, these modes typically shorten metal-metal bond distances as temperature increases through relative rotations of rigid (or almost rigid) polyhedral groups, or kinking of oxygen-metal-oxygen bond networks. The RUM model of NTE successfully accounts for the thermal behavior of the canonical NTE material ZrW$_2$O$_8$ \cite{tucker2005negative,alexandra1996origin,mary1996negative}, as well as a number of zeolites \cite{hammonds1998rigid,lightfoot2001widespread} and Prussian blue materials \cite{goodwin2005guest}. A second essential ingredient for the occurrence of NTE appears to be the existence of low-frequency RUMs with large, negative Gr\"{u}neisen parameters, that is, modes with frequencies that \emph{decrease} with decreasing volume. In fact, negative Gr\"{u}neisen parameters are sometimes claimed to be prerequisites for NTE behavior \cite{yamamura2009characteristic, mittal2017phonons}. Indeed, as far as we know, the only exceptions are the elemental metals Zn \cite{mccammon1965thermal}, Cd \cite{mccammon1965thermal}, As \cite{white1972thermal}, and Sb \cite{white1972thermal}, all of which exhibit only uniaxial, rather than volumetric, NTE.

Given the apparent importance of RUMs and phonons with negative Gr\"{u}neisen parameters, it is curiously rare for ABO$_3$ perovskites to undergo NTE. Most perovskites undergo one or more structural phase transitions involving phonons with strong RUM-like character (typically rotations of the BO$_6$ octahedra), and these types of modes are associated with negative Gr\"{u}neisen parameters in, for example, $\beta$-cristobalite, $\beta$-quartz, and ZrW$_2$O$_8$ \cite{dove2016negative}. In contrast, recent work \cite{ablitt17,senn2015negative} has shown that (uniaxial) NTE is common in \emph{layered} perovskites, such as Ruddlesden-Popper phases, because the combined effects of layering and rotations of the BO$_6$ octahedra enhance their elastic anisotropy compared to bulk perovskites; this enhanced elastic anisotropy appears to be the origin of uniaxial NTE in layered perovskites. Of course, bulk perovskites may exhibit elastic anisotropy, as Refs. \onlinecite{ablitt17} and \onlinecite{senn2015negative} acknowledge, however the magnitude of the anisotropy is generally not large enough to induce NTE.

In this Letter, we use the ferroelectric perovskite PbTiO$_3$ to demonstrate that neither RUMs nor phonons with negative Gr\"{u}neisen parameters are necessary (or sufficient) in order for a material to undergo NTE. A key discovery we make is that in non-cubic materials, NTE cannot be predicted based on either the signs or magnitudes of an individual Gr\"{u}neisen parameter or elastic constant by itself, as is commonly assumed. In these systems, the thermal expansion along a given axis is coupled to multiple Gr\"{u}neisen parameters through multiple \emph{independent} elastic constants. We use theory and first-principles calculations to elucidate the microscopic mechanism of NTE in PbTiO$_3$ and show that few of the modes critical to driving NTE have negative Gr\"{u}neisen parameters and that they are not RUMs, even though PbTiO$_3$ contains many RUM-like modes \cite{hammonds1998rigid}. We then connect the physical mechanism of NTE in PbTiO$_3$ to its electronic structure, and show that its elastic properties are dominated by the stereochemical activity of the Pb$^{2+}$ 6$s^2$6$p^0$ lone electron pair. Our results are striking because they suggest that PbTiO$_3$ is unique among well-studied NTE materials, and appears to be the only material that exhibits \emph{volumetric} NTE well above room temperature, and with positive Gr\"{u}neisen parameters along all unique crystallographic axes. These results bring clarity to the driving mechanism of NTE in PbTiO$_3$, and broaden the search for new NTE systems to include materials overlooked in the past due to a lack of large, negative Gr\"{u}neisen parameters.

\begin{figure}
\centering
\includegraphics[width=7.8cm]{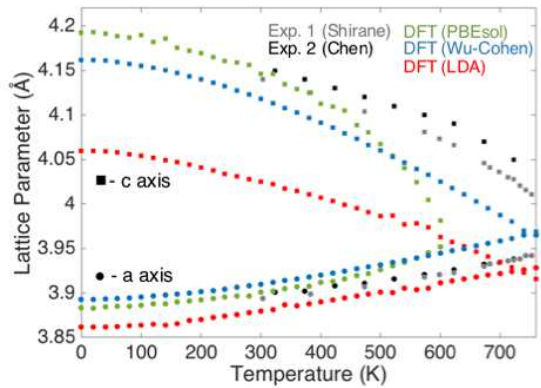}
\caption{(Color online) Lattice parameters (\textit{a} filled circles, \textit{c} filled squares) as a function of temperature from first-principles calculations (colored icons) and from experiment (black and gray icons \cite{shirane1951phase,chen2005structure}).}
\label{lattice}
\end{figure}

PbTiO$_3$ is cubic at high temperatures but undergoes a phase transition at 760 K to a tetragonal (\textit{P4mm}) ferroelectric phase; this phase exhibits volumetric NTE down to approximately room temperature (though only the $c$ axis decreases with temperature while the $a$ axes increase, the net effect is a decrease in volume). We used density functional theory, as implemented in Quantum Espresso with GBRV pseudopotentials \cite{giannozzi2009quantum, garrity2014pseudopotentials} (see Supplementary Information (SI) for methods and convergence critera) to calculate the evolution of lattice parameters as a function of temperature in the quasiharmonic approximation (QHA). Figure \ref{lattice} compares our results obtained from three functionals: LDA \cite{kohn1965self}, PBEsol \cite{ernzerhof1999assessment} and Wu-Cohen (WC) \cite{WC2006}. As expected, all three functionals slightly underestimate the lattice parameters, however each qualitatively reproduces the experimental trend of a shrinking $c$ axis and lengthening $a$ axis as temperature increases. Our calculations reproduce the experimentally observed volumetric NTE, and additional comparisons between data from our first-principles calculations and available experimental data for finite-temperature phonon dynamics and elastic constants indicate that the QHA is a justified approximation for this system (see SI). We report all remaining results for the WC functional only, since it best captures the structural properties of PbTiO$_3$. With this functional, we calculate a volumetric expansion coefficient $\alpha_v$ of $-2.29 \times 10^{-5}$ K$^{-1}$ between 500 K and 700 K, which compares favorably with $\alpha_v=-1.8 \times 10^{-5}$ K$^{-1}$ and $-1.99 \times 10^{-5}$ K$^{-1}$  from  \cite{shirane1951phase} and \cite{chen2005structure}, respectively.   

We now turn to elucidating the microscopic mechanism of NTE in PbTiO$_3$. We first calculated the Gr\"{u}neisen parameters for the equilibrium structure of PbTiO$_3$ at 300 K. Since PbTiO$_3$ is not cubic, the Gr\"{u}neisen parameter $\gamma$ has a tensor form,
\begin{equation}
\gamma^{ij}_{s,\mathbf{k}}\equiv-\frac{1}{\omega_{s,\mathbf{k}}} \frac{\partial \omega_{s,\mathbf{k}}}{\partial \varepsilon_{ij}},
\label{modeGrunTens}
\end{equation}
where $\omega_{s,\mathbf{k}}$ is the frequency of mode $s$ at wavevector $\mathbf{k}$, $\varepsilon_{ij}$ is a strain consistent with crystal symmetry, and $i$ and $j$ are Cartesian directions. We calculated $\gamma^{ij}_{s,\mathbf{k}}$ using a central difference, which required three separate calculations of the full dispersion curve.  The quantity usually referred to as the `bulk' Gr\"{u}neisen parameter is then defined as \cite{barron1967analysis}, 
\begin{equation}
\gamma_{\mathrm{bulk}}^{ij} = \frac{\sum_{s,\mathbf{k}}\gamma_{s,\mathbf{k}}^{ij}c_{s,\mathbf{k}}}{\sum_{s,\mathbf{k}}c_{s,\mathbf{k}}},
\label{gbulk}
\end{equation}
where $c_{s,\mathbf{k}}$ is the mode specific heat at constant configuration. We note that the uniaxial stress perturbation method \cite{romao2017anisotropic} could be suited to high-throughput studies to screen for NTE along particular axes, and to provide a quick alternative method to compare with results obtained using the strain perturbative definitions as in Eq. (\ref{modeGrunTens}).

\begin{figure}
\includegraphics[width=8.6cm]{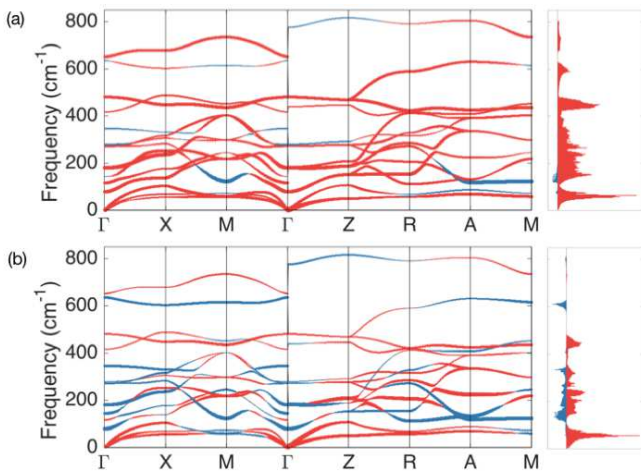}
\caption{(Color online) {\bf (a)} Phonon dispersion curve for PbTiO$_3$ at 300 K with band thickness proportional to the magnitude of $\gamma^{a}_{s,\mathbf{k}}$ for each mode, and color corresponding to sign of $\gamma^{a}_{s,\mathbf{k}}$ (red positive, blue negative). To the right is the sum of $\gamma^{a}_{s,\mathbf{k}}c_{s,\mathbf{k}}$ across entire Brillouin zone for each energy level at 300 K, positive (red) and negative (blue) contributions plotted separately. {\bf (b)} Same, corresponding to $\gamma^{c}_{s,\mathbf{k}}$ }
\label{grunPlot}
\end{figure}

\begin{figure}
\includegraphics[width=8.6cm]{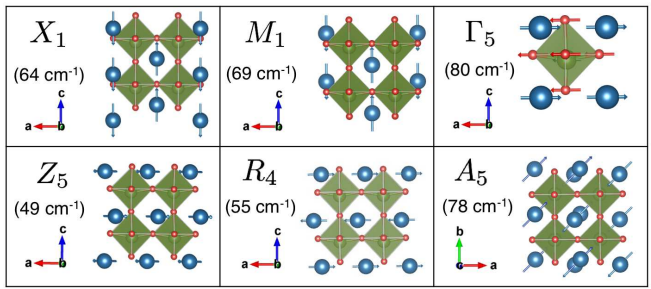}
\caption{(Color online) Representative set of distortions associated with low-frequency ($\omega <$ 100 cm$^{-1}$) phonons for the indicated irreducible representation of each high symmetry point. Note the absence of RUMs and dominance of Pb motion. See SI for additional modes.}
\label{distPlot}
\end{figure}

Figure~\ref{grunPlot} shows phonon dispersion curves for PbTiO$_3$ at 300 K with the magnitude and sign of $\gamma^{a}_{s,\mathbf{k}} \equiv \gamma^{11}_{s,\mathbf{k}}  $ and $\gamma^{c}_{s,\mathbf{k}} \equiv \gamma^{33}_{s,\mathbf{k}} $ represented by the thickness and color of the band, respectively. Despite the presence of NTE, both bulk Gr\"uneisen parameters are positive,  $\gamma^a_{\mathrm{bulk}}$ = 1.42 and $\gamma^c_{\mathrm{bulk}}$ = 0.40. The density of states in Figure~\ref{grunPlot} indicates that the individual phonon modes contributing most strongly to $\gamma^a_{\mathrm{bulk}}$ and $\gamma^c_{\mathrm{bulk}}$ are low frequency modes with \emph{positive} Gr\"{u}neisen parameters, in contrast with the usual expectation that modes with large, negative mode Gr\"uneisen parameters drive NTE. As shown in the Supplementary Information, if we force all modes below 100 cm$^{-1}$ to be perfectly harmonic by keeping their frequencies constant with temperature, while allowing the frequencies of all other modes to change, then NTE behavior is completely suppressed. Hence, these low-frequency modes appear to be the primary drivers of NTE and of the positive bulk Gr\"{u}neisen parameters along both the $a$ and $c$ axes. Additionally, Figure \ref{distPlot} shows that \emph{none} of the distortions associated with the critical low-frequency modes are RUMs; they are  instead dominated by translational Pb motion. \cite{ghosez99} Our work so far raises two questions: how do we account for NTE in PbTiO$_3$, given the positive Gr\"{u}neisen parameters, and do electronic effects play a role in NTE in this material?

Although large, negative mode Gr\"uneisen parameters have been associated with NTE in the literature, this relationship need not hold for anisotropic systems. In cubic materials, the bulk Gr\"{u}neisen parameter is scalar and the coefficient of volumetric thermal expansion can be expressed as \cite{ashcroft2005solid}
\begin{equation}
\alpha_v=\frac{\gamma_{\mathrm{bulk}}C_{\eta}}{3B},
\label{coefTherm}
\end{equation}
where $B$ is the bulk modulus and $C_{\eta}$ is the specific heat at constant configuration. The bulk Gr\"{u}neisen parameter $\gamma_{\mathrm{bulk}}$ is still defined as in Eq. (\ref{gbulk}), however the individual mode Gr\"{u}neisen parameters are reduced to simple volume derivatives,
\begin{equation}
\gamma_{s,\mathbf{k}}\equiv-\frac{V}{\omega_{s,\mathbf{k}}}\frac{d\omega_{s,\mathbf{k}}}{dV}.
\label{modeGrunVol}
\end{equation}
Looking at Eq. (\ref{coefTherm}), $\gamma_{\mathrm{bulk}}$ is the only term on the right-hand side that can be negative. Hence, a negative $\alpha_v$ in {\it cubic} systems is impossible without a negative $\gamma_{\mathrm{bulk}}$, which must itself arise from large, negative mode Gr\"{u}neisen parameters.

Anisotropic materials have additional structural degrees of freedom that complicate the relationship between the coefficient of thermal expansion, Gr\"{u}neisen parameters, and elastic constants. For example, in tetragonal systems like PbTiO$_3$, the coefficients of thermal expansion along $a$ ($\alpha_{a}$) and $c$ ($\alpha_{c}$) are defined as \cite{wallace,munn1972role}:
 \begin{equation}
\alpha_{a}=(C_{\eta}/V)\bigg[(S_{11}+S_{12})\gamma^a_{\mathrm{bulk}}+S_{13}\gamma^c_{\mathrm{bulk}} \bigg],
\label{alpha2}
\end{equation}
and
\begin{equation}
\alpha_{c}=(C_{\eta}/V)\bigg[2S_{13}\gamma^{a}_{\mathrm{bulk}}+S_{33}\gamma^c_{\mathrm{bulk}}\bigg], \newline
\label{alpha1}
\end{equation}
where $S_{ij}$ is the $ij^{th}$ component of the elastic compliance tensor $\mathbf{S}$ in Voigt notation. Since $\alpha_v \equiv 2\alpha_{a}+\alpha_{c}$ \cite{wallace}, in order for $\alpha_v$ to be negative, we require
\begin{equation}
 2(S_{11}+S_{12}+S_{13})\gamma^{a}_{\mathrm{bulk}}+(S_{33}+2S_{13})\gamma^c_{\mathrm{bulk}}  <0.
 \label{NTEcondition}
 \end{equation}
Eq. (\ref{NTEcondition}) shows that negative bulk Gr\"{u}neisen parameters are not a prerequisite for NTE in anisotropic materials. The S$_{ij}$ can be positive or negative, and the relative sign and magnitude of these terms constrain the signs of $\gamma^a_{\mathrm{bulk}}$ and $\gamma^c_{\mathrm{bulk}}$ required for NTE. Without knowledge of the full compliance tensor, \emph{no definitive statements about how the signs or magnitudes of $\gamma^a_{\mathrm{bulk}}$ and $\gamma^c_{\mathrm{bulk}}$ relate to $\alpha_v$ can be made}. However, the dependence of $\alpha_v$ on $\gamma^c_{\mathrm{bulk}}$ and $\gamma^a_{\mathrm{bulk}}$ can be determined directly if S$_{ij}$ is known.

Table \ref{compliance} shows selected elements of the compliance tensor for the tetragonal phase of PbTiO$_3$ from first principles calculations, which compare well with experiment \cite{kalinichev1997elastic} (note the large magnitude of $S_{13}$, previously correlated \cite{goodwin2008colossal,ablitt17} with NTE behavior). The $a$-axis expands as temperature rises, and so $\alpha_a$ is positive. From Eq. (\ref{alpha2}), since $S_{11} + S_{12} > 0$ for PbTiO$_3$, $\gamma^a_{\mathrm{bulk}}$ must be large and positive in order for $\alpha_a > 0$. Since positive Gr\"{u}neisen parameters are known to be associated with positive thermal expansion, our results so far are unsurprising. However, we will now argue that it is the tendency of PbTiO$_3$ to expand along $a$ with increasing temperature that \emph{drives} the system to shrink along $c$, giving rise to volumetric NTE.

The key to understanding NTE in PbTiO$_3$ is to examine the relationship between the Gr\"{u}neisen parameters, and applied stress along the $a$ axis and induced strain along the $c$ axis. Since the $c$ axis contracts as temperature increases, $\alpha_c < 0$. The conventional wisdom suggests that this requires a negative $\gamma^c_{\mathrm{bulk}}$, while we know that $\gamma^c_{\mathrm{bulk}}$ is actually positive in this system. From Eq. (\ref{alpha1}), since $S_{33}$ is also positive, the only way for $\alpha_c$ to be negative is for the product $S_{13}\gamma^a_{\mathrm{bulk}}$ to be negative -- indeed the case for PbTiO$_3$, as shown in Table \ref{compliance}.  Physically, a large, negative $S_{13}$ in tetragonal systems functions in a similar way to a large, positive Poisson's ratio in cubic systems. \emph{It is this temperature-induced negative strain that drives $\alpha_c$ to be negative, and gives rise to NTE in PbTiO$_3$.} 

If $S_{13}\gamma^a_{\mathrm{bulk}}$ was positive, or even slightly less negative, then it would be energetically favorable for the $c$ axis to expand with increasing temperature. Hence, the NTE is not due to a $c$ axis driven to contract by a negative $\gamma^c_{\mathrm{bulk}}$, but \emph{in spite of} a positive $\gamma^c_{\mathrm{bulk}}$. We illustrate this by rewriting equations \ref{alpha2} and \ref{alpha1} in matrix form:

\begin{equation}
\begin{bmatrix}
\alpha_a \\
\alpha_a \\
\alpha_c \end{bmatrix} =
\begin{bmatrix}
S_{11} & S_{12} & S_{13}\\ 
S_{12} & S_{11} & S_{13}\\
S_{13} & S_{13} & S_{33}\end{bmatrix} 
\begin{bmatrix} 
\frac{C_{\eta}\gamma^a_{\mathrm{bulk}}}{V}  \\
\frac{C_{\eta}\gamma^a_{\mathrm{bulk}}}{V} \\
\frac{C_{\eta}\gamma^c_{\mathrm{bulk}}}{V} \end{bmatrix}.
\end{equation}

The column vector on the right has units of pressure/Kelvin -- a temperature-induced stress -- linked through the compliance tensor to the column vector on the left, which has units of strain/Kelvin. In this arrangement, $\gamma^a_{\mathrm{bulk}}$ and $\gamma^c_{\mathrm{bulk}}$ define the magnitude of a thermal stress normal to the $a$ and $c$ axes, respectively, and are linked to thermal strain through $\mathbf{S}$. Just as a mechanical stress state involving positive stress along both axes could result in negative strain along $c$ due to coupling through $S_{13}$, thermal stress driven by positive $\gamma^a_{\mathrm{bulk}}$ and $\gamma^c_{\mathrm{bulk}}$ could result in $\alpha_c<0$, where the sign of $\alpha_c$ depends on a careful balance between \emph{all} of $S_{ij}$, $\gamma^a_{\mathrm{bulk}}$ and $\gamma^c_{\mathrm{bulk}}$. 

\begin{table}
\centering
\caption{\label{compliance}Selected elements of the compliance tensor for PbTiO$_3$ and SnTiO$_3$ from first-principles calculations in units of $10^{-3}$ GPa$^{-1}$.}
\begin{tabular}{ccccc}
 & $S_{11}$ & $S_{12}$ & $S_{13}$ & $S_{33}$  \\ \hline
PbTiO$_3$  & 7.44  & 0.49  & -11.94  & 55.69 \\ \hline
SnTiO$_3$  & 7.79  & -1.45  & -6.83  & 31.36 \\ \hline
\end{tabular}
\end{table}

Understanding that NTE in PbTiO$_3$ arises from a careful balance between $S_{ij}$, $\gamma^a_{\mathrm{bulk}}$, and $\gamma^c_{\mathrm{bulk}}$ terms, we can explore the origin of the sign and magnitude of these parameters through investigating a similar ABO$_3$ perovskite, SnTiO$_3$. Previous {\it ab initio} studies predict that SnTiO$_3$ also adopts a $P4mm$ ground state structure at 0 K  \cite{parker2011first}. Though synthesis of bulk SnTiO$_3$ has proven difficult \cite{matar2009first, konishi2002possible} and few reliable experimental measurements of physical properties exist, it still provides a useful comparison to illuminate the origin of NTE in PbTiO$_3$. We find that within the QHA, SnTiO$_3$ exhibits repressed NTE behavior compared to PbTiO$_3$, and positive or near-zero volumetric NTE at every temperature (see SI). 

The behavior of SnTiO$_3$ can be linked to terms in the compliance matrix (Table \ref{compliance}) and bulk Gr\"uneisen tensor ($\gamma^c_{\mathrm{bulk}} = 7.72$, $\gamma^a_{\mathrm{bulk}} = 3.42$). SnTiO$_3$ has similar compliance as PbTiO$_3$ along the $a$ axis, but is much stiffer (lower compliance) along $c$ and exhibits much less coupling between stress along $a$ and strain along $c$ (a lower $S_{13}$). SnTiO$_3$ exhibits a slightly more positive $\gamma^a_{\mathrm{bulk}}$ at 300 K than PbTiO$_3$ ($3.42$ to $1.42$), but a $\gamma^c_{\mathrm{bulk}}$ over an order of magnitude larger ($7.72$ to $0.40$). Referring to Eqs. (\ref{NTEcondition}) and (\ref{alpha1}), we can confirm that these differences should result in a system where NTE is suppressed compared to PbTiO$_3$. 

As $\gamma^a_{\mathrm{bulk}}$ and $\gamma^c_{\mathrm{bulk}}$ are vibrational, dynamic properties, the difference between these values in each material could be due to either the mass difference between Sn and Pb (118.71 a.u. and 207.2 a.u., respectively) or from differences in electronic structure, primarily the tendency of the A-site cation to form stereochemically active lone pairs. We rule out mass difference as the driving effect by performing a computational ``experiment" in which the QHA is performed for PbTiO$_3$, but with the mass of Pb changed to that of Sn (see SI). PbTiO$_3$ then exhibits lattice parameters as a function of temperature nearly identical to those for the original system in Figure \ref{lattice}, and a 500-700 K $\alpha_v$ of $-2.38 \times 10^{-5} K^{-1}$, which is within 5\% of $\alpha_v$ for the original system. Though this experiment is not physically realizable, the results show that the absence of NTE in SnTiO$_3$ {\it cannot} be due to the mass difference between Pb and Sn alone.

Previous studies \cite{walsh2011stereochemistry, stoltzfus2007structure, pitike2015first, waghmare2003first} have shown that the n$s^2$n$p^0$ lone electron pair on Pb and Sn is stereochemically active and is responsible for driving the symmetry-breaking distortion that lowers the symmetry of each material from cubic to tetragonal. The Pb/Sn-$s$-anion-$p$ interaction at the top of the valence band is actually antibonding and therefore energetically destabilizing. These states could be stabilized by mixing with the unfilled Pb/Sn-$p$ states at the bottom of the conduction band, but such mixing is forbidden by symmetry in the cubic perovskite structure. This electronic instability (known as a pseudo- or second-order Jahn-Teller distortion) manifests in the lattice as a structural phase transition: the ferroelectric transition lowers the atomic site symmetries such that Pb/Sn-$s$-anion-$p$-Pb/Sn-$p$ mixing is allowed. The resulting localized, non-bonding state at the top of the valence band is the lone pair.

The closer the cation-$s$ and anion-$p$ states are in energy, the more cation-$s$ character will be present in the anti-bonding states, and the greater the stabilization gained from hybridization between these anti-bonding states and the unfilled cation-$p$ states. Since the Sn 5$s$ states are closer in energy to the O 2$p$ states than the Pb $6s$ states are \cite{walsh2011stereochemistry}, SnTiO$_3$ exhibits a stronger tendency towards a structural distortion than PbTiO$_3$. In fact, Crystal Orbital Hamiltonian Population (COHP) analysis \cite{dronskowski1993crystal,deringer2011crystal,maintz2013analytic,maintz2016lobster} shows that while the Jahn-Teller distortion makes the Pb-$s$ -- O-$p$ interaction slightly less antibonding, in the case of SnTiO$_3$, it brings the Sn-$s$ -- O-$p$ interaction from antibonding in the cubic phase to \emph{bonding} in the tetragonal phase. Since these states are filled, there is a larger energy gain associated with the Jahn-Teller distortion in SnTiO$_3$ compared with PbTiO$_3$, and a larger corresponding structural distortion.

Next, we link the size of the structural distortion in SnTiO$_3$ to $S_{13}$ and its NTE behavior. It is well known in the ferroelectrics literature \cite{ederer05,lee07} that $c/a$ in tetragonal BaTiO$_3$ can be enhanced signficantly with epitaxial strain, thereby signficantly enhancing ferroelectric polarization. In contrast, $c/a$ (and therefore polarization) in PbTiO$_3$ is quite insensitive to epitaxial strain. This behavior is attributed to the fact that PbTiO$_3$ in the tetragonal phase is already very structurally distorted, and hence resists further distortions. Since $c/a$ for SnTiO$_3$ (1.18) is larger than PbTiO$_3$ (1.08), we predict that $S_{13}$ for SnTiO$_3$ is less negative than for PbTiO$_3$ because SnTiO$_3$ is \emph{even more} resistant to further structural distortions than PbTiO$_3$. We tested this hypothesis by calculating selected elements of the compliance tensor for a hypothetical SnTiO$_3$ structure with a $c/a$ ratio set to that of PbTiO$_3$. Compared with SnTiO$_3$ with its equilibrium $c/a$, this hypothetical structure has a more negative $S_{13}$, a larger (more positive) $S_{33}$, and a $\gamma^c_{bulk}$ of -1.71. Inserting these values of $S_{ij}$ and Gr\"{u}neisen parameters into Eq. (\ref{alpha2}) and (\ref{alpha1}), we find $\alpha_a$ = 7.5 ($\times$10$^{-4}$K$^{-1}$) and $\alpha_c$ = -26.0 ($\times$10$^{-4}$K$^{-1}$). Since $\alpha_v\equiv 2 \alpha_a + \alpha_c$, this hypothetical SnTiO$_3$ structure should exhibit volumetric negative thermal expansion. Hence, the electronic structure of the material is important in determining the size of the structural distortion away from the high-symmetry phase. However, in the distorted structure, it is primarily the \emph{shape} of the unit cell that determines the elastic properties and Gr\"{u}neisen parameters (see SI for further details). 

Our results suggest that revisiting anisotropic systems with positive bulk Gr\"{u}neisen parameters may be a promising direction in the search for new NTE materials. Since the number of unique indices of $\mathbf{S}$ and $\gamma^{ij}_{\mathrm{bulk}}$ are larger for Bravais lattices of lower symmetry, the number of degrees of freedom in expressions like Eq. (\ref{NTEcondition}) increases, providing more pathways for NTE. We have also shown that, critically, as volumetric NTE arises from the product of multiple $\gamma^{ij}_{\mathrm{bulk}}$ and $S_{ij}$, a search for NTE materials that prioritizes \emph{either} large magnitudes \emph{or} particular signs of certain indices of these tensors will miss favorable candidates. Additionally, in light of previous work \cite{nielsen2013lone} suggesting that the presence of lone pair electrons reduces thermal conductivity, exploring the link between phonon modes driving NTE in PbTiO$_3$ and those driving thermal expansion could yield new insights with respect to mode-level control of thermal properties.

\begin{acknowledgments}
Initial work on this project by ETR was supported by NASA Space Technology Research Fellowship award number NNX13AL39H, and subsequently by the National Science Foundation under award number DMR-1550347. NAB was supported by the National Science Foundation under award number DMR-1550347. This work used the Extreme Science and Engineering Discovery Environment (XSEDE) (through allocation DMR-160052), which is supported by National Science Foundation grant number ACI-1548562. Additional high-performance computing resources were provided by the Cornell Center for Advanced Computing.  We acknowledge helpful discussions with Eric Toberer, Craig Fennie, and Guru Khalsa. 
\end{acknowledgments}

\bibliography{citations}

\end{document}